# Comparison Among Coaxial Microcalorimeter Models

L. Brunetti, L. Oberto, M. Sellone, and E. Vremera

*Abstract*—Thermoelectric power sensors used as power transfer standards are promising devices for further enhancements of the microcalorimetric technique in the high frequency field. A coaxial microcalorimeter has been studied, based on thermoelectric power sensors, at Istituto Nazionale di Ricerca Metrologica (INRIM). In the literature, several models have been proposed, considering different calibration processes and error sources. Hereby we analyze these models in terms of total uncertainty for the 3.5 mm coaxial line case between 10 MHz and 26.5 GHz. Merits and limits of the models are highlighted.

## I. INTRODUCTION

Since its origin, dating back to the late 1950s, the microcalorimeter technique is the best solution for realizing *primary power standards* at high frequency (HF) [1], [2]. Its main goal is tracing these standards to the direct current (dc) standard through the determination of the power sensor effective efficiency ($\eta_e$) versus the frequency.

In general, the microcalorimeter technique is based on transfer standards of bolometric type, i.e. power sensors, that measure the HF power by means of dc power substitution [3]. Theory and technique have mainly concerned waveguide systems because they are easier to realize and study, even at millimeter wavelengths [4]. Progress in the development of coaxial power sensors has made the broadband coaxial Microcalorimeter more attractive. Even though bolometric power sensors are still important for coaxial transmission line measurements, they may be substituted by indirect heating thermoelectric sensors with advantages and drawbacks, on bolometric detectors, that have repeatedly been reported in literature [5], even of the same authors [6], [7]. It must be highlighted, however, that only dc-coupled sensors can operate with the dc power substitution principle. It is the case of the indirect heating thermocouples that, indeed, have successfully been used as transfer standards even in a key-comparison for HF power in coaxial line up to 26 GHz [8], where substitution was suggested at 1 kHz to avoid the bias error due to contact thermo-voltages. Since several years, INRIM is operating with a coaxial microcalorimeter that is adapted for modified commercial thermoelectric power sensors [7]. The system has been improved several times both in its hardware components and in the software.

In this paper we present the related mathematical models that we were able to propose together to a comparison of their accuracy. We consider the case of a 3.5 mm transmission line between 0.01 and 26.5 GHz because of the significant reference available under these conditions [8].

## II. INRIM COAXIAL MICROCALORIMETER

The INRIM system is based on a dry thermostat [6], [7], whose triple-wall measurement chamber is thermally stabilized via Peltier cells acting on the intermediate wall. The other two walls work only as passive thermal shields with different thermal characteristics. The thermostat has been designed to operate into a conditioned shielded room at (23.0 ± 0.3) °C and (50 ± 5) % of relative humidity.

This solution allows to obtain a thermal stability that can be given in terms of temperature fluctuations measured on the active wall, that is (25.00 ± 0.01) °C. The measurement chamber should have a thermal stability one order better.

After an experience of many years on single line insets, the twin-line configuration of Fig. 1 has been adopted, because it is more effective in the suppression of the thermal noise that still bypass the thermal shields. The system can operate both with bolometric power sensors and thermoelectric devices. However, measurement procedures and related models are hereby relevant to thermoelectric transfer standards only.

## III. THEORY AND MATHEMATICAL MODELS

For thermoelectric power sensors, an *effective efficiency* definition does not exist that is unconditionally accepted and reported in the technical literature as for the bolometric case [9]. The ratio between a low frequency (LF) reference power and the HF power that produces the same sensor output

Manuscript received June 5, 2008.
L. Brunetti is with the Istituto Nazionale di Ricerca Metrologica (INRIM), Electromagnetism Division, strada delle Cacce 91, 10135 Torino, Italy (corresponding author to provide phone: +39(0)113919323; fax: +39(0)113919259; e-mail: l.brunetti@inrim.it).
L. Oberto and M. Sellone are with the Istituto Nazionale di Ricerca Metrologica (INRIM), (e-mail: l.oberto@inrim.it and m.sellone@inrim.it).
E. Vremera is with the Faculty of Electrical Engineering, Technical University of Iasi, Bld. Dimitrie Mangeron 53, Iasi, 700050 Romania (e-mail: evremera@ee.tuiasi.ro).

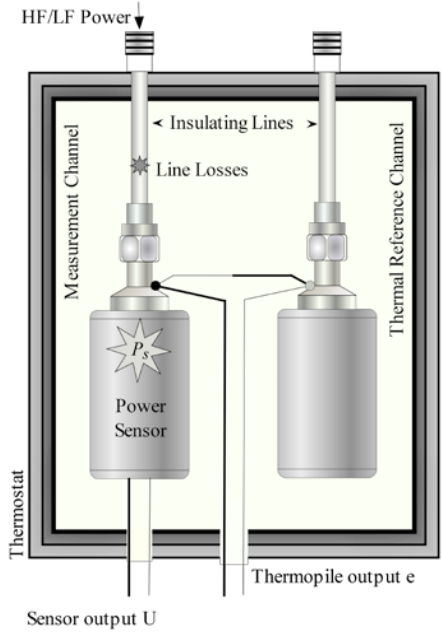

Fig. 1. Twin-type coaxial microcalorimeter scheme. $P_S$ represents the total power dissipated into the sensor.

($U=const$) was adopted in [8].

This was an opportunity definition suggested by the protocol of the comparison to avoid misunderstanding and to assure uniformity among the results. Here we replace the axiomatic definition of *effective efficiency* previously given with the more general and intuitive expression:

$$\eta_e = \frac{P_M}{P_M + P_X}, \qquad (1)$$

where $P_M$ is the power measured by the sensor, i.e. the power converted to the sensor dc-output $U$ and $P_X$ is the power loss in sensor mount at the same frequency.

Supplying power to the microcalorimeter, the system asymptotic response is given by:

$$e = \alpha R (K_A P_S + K_B P_L), \qquad (2)$$

where $\alpha$ and $R$ are conversion coefficients, $P_S$ is the total power dissipated into the sensor and $P_L$ is the power dissipated into the feeding line. $K_A$ and $K_B$ are power separation coefficients dependent on thermal characteristics of the microcalorimeter inset [6].

Our measurand $\eta_e$ is, however, embedded in (2), that is, basically, an equation having four unknowns. To obtain it in form (1), we have to re-write (2) for two different frequencies and then rationing the results.

Introducing the thermopile asymptotic voltages $e_1$ related to a HF power and $e_2$ related to the system response when a LF power is substituted to the first, we obtain the following relation:

$$\frac{e_1}{e_2} = \frac{P_{1M} + P_{1X}}{P_{1M}} \frac{U_1}{U_2} \left[ \frac{P_{2M}}{P_{2M} + P_{2X}} \right] \left[ \frac{1 + \frac{K_B}{K_A} \frac{P_{1L}}{P_{1S}}}{1 + \frac{K_B}{K_A} \frac{P_{2L}}{P_{2S}}} \right], \qquad (3)$$

where $U_1$, $U_2$ are the sensor dc response to HF and LF, respectively. Index 1 refers to HF power, while 2 to a specific LF power, named reference power, that is 1 kHz for us and in the following. Observing (3), we can recognize the $\eta_e$ definition given by (1), if the power substitution is made under the condition $U_1 = U_2$. However, square bracket terms are unknowns that must be somehow determined. The left one has the form of an effective efficiency at 1 kHz, while the other has the following explicit expression:

$$g = \frac{\left( 1 + \frac{K_B}{K_A} \frac{(1-|S_{21}|^2)(1+|S_{21}|^2|\Gamma_S|^2)}{|S_{21}|^2 (1-|\Gamma_S|^2)} \right)\bigg|_{HF}}{\left( 1 + \frac{K_B}{K_A} \frac{(1-|S_{21}|^2)(1+|S_{21}|^2|\Gamma_S|^2)}{|S_{21}|^2 (1-|\Gamma_S|^2)} \right)\bigg|_{1kHz}}, \qquad (4)$$

where $S_{21}$ and $\Gamma_S$ are the transmission parameter of the feeding line and the reflection coefficient of the power sensor, respectively. The effective efficiency at 1 kHz cannot be found by this method so it has to be considered as a reference value. Even if enough experimental evidence exists for considering $\eta_e = 1$ at 1 kHz, we have to recognize that the effective efficiency obtained with the microcalorimeter technique is not an absolute value. Instead, it is the ratio between the effective efficiency of the sensor under calibration at the frequency of interest and its reference efficiency.

As there are still no official guidelines in the determination of $\eta_e$ for thermocouple based power sensors, it is very important to establish a common reference frequency to be used by NMIs, to avoid incomparable results. In order to maintain the compatibility of newer measurements with the results obtained in [8], we operate with the reference frequency of 1 kHz for the realization of primary HF power standards based on thermoelectric devices of this kind.

Nevertheless, $g$ has to be determined through a calorimetric experiment, mainly because $K_A$ and $K_B$ cannot be measured by a network analyzer as it is, instead, done for the scattering parameters $S_{21}$ and $\Gamma_S$. $K_A$ and $K_B$ are, moreover, scalar quantity related to the thermal conductance and thermal capacity of the microcalorimeter inset through a very complicated and not well determinable thermodynamical model.

For going further, it is necessary to invert Eq. (3), assuming $\eta_e = 1$ at 1 kHz, which can be rewritten as follows:

$$\frac{e_1}{e_2} = \frac{1}{\eta_e}\left(\frac{U_1}{U_2}\right)g \qquad (5)$$

This operation requires a power standard calculable or measurable with an independent method. This is true for the laboratories that attempt to realize a power standard in completely independent manner from others. A measurable $\eta_e$ can be obtained by short-circuiting the sensor input connector appropriately [7]. Up to now, we have to assume model (3) to be ineffective for the state of the art of the thermoelectric sensors. In fact, when the sensor under calibration is substituted with the short circuited one, there is no output signal $U$, so the ratio $U_1/U_2$ cannot be calculated. Conversely, model (5) could be very useful to update the microcalorimeter calibration constant $g$ and, indirectly, the ratio $K_B/K_A$, if the reference values of relevant international comparisons are available. In this case, (5) can be easily reversed by using previous data without measurements or hardware operations.

Working around the mentioned problems, we have elaborated a simplified model that proved to be effective and reliable [7]. Under the assumption of negligible feeding line losses at 1 kHz and with a power control loop able to maintain constant the sensor output $U$ at each power substitution, the measurand $\eta_e$ is also given by:

$$\eta_e = \frac{e_2}{e_1 - e_{1SC}}, \qquad (6)$$

where $e_1$, $e_2$ are still the asymptotic responses of the microcalorimeter thermopile at the HF and LF power, respectively, as in (3), while $e_{1SC}$ is the response when half of the HF power generating $e_1$ is supplied to a totally reflective load [6]. It is a corrective term related to feeding line losses. The details for operating correctly are reported in [7].

The assumption done is equivalent to eliminate an unknown, i.e. $P_L$ at 1 kHz, from the general problem. It is supported by auxiliary measurements on the feeding line. Indeed, both at dc and at 1 kHz, the ohmic resistance of the microcalorimeter inset resulted <0.01Ω, a value that does not introduce a significant perturbation.

In model (6), we suppose that the short circuit used for calibration is perfect. In practice, to realize a reflecting short thermally equivalent to the sensor under calibration, we modify (although reversibly) the sensor input port or the microcalorimeter feeding line output port. Because the reflection coefficient $\Gamma_{SC}$ of the resulting device has normally a significant performance degradation versus frequency, as Fig. 2 shows, Eq. (6) should be corrected including another term related to the short circuit losses at HF. These ones are proportional to $(1-|\Gamma_{SC}|^2)$, while an effective efficiency equal to $|\Gamma_{SC}|^2$ can be associated to a short circuited power sensor [10]. This can be mathematically proved but also intuitively explained by considering that the system can only recognize the losses in its thermal load, but not the real electrical

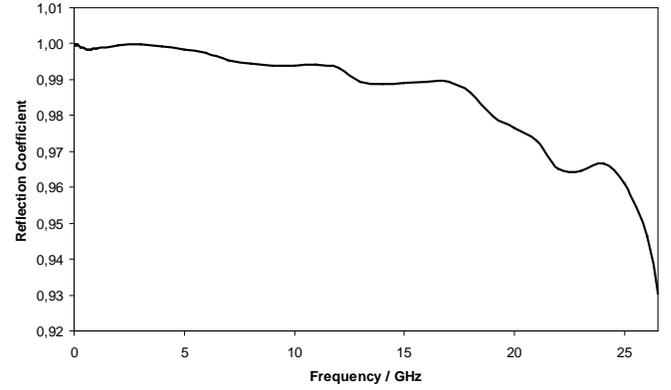

Fig. 2. Behavior versus frequency of the reflecting device used for the system calibration.

characteristic of it.

Applying the procedure detailed in [7] to derive Eq. (6), including also the short losses, we get the following model:

$$\eta_e = \frac{e_2}{e_1 + e_{2SC}|\Gamma_{SC}|^{-2} - e_{1SC}} \qquad (7)$$

in which the term $e_{2SC}$ is the microcalorimeter thermopile response when the shorted feeding line is supplied with the reference power at 1 kHz. Moreover, Eq. (7) can be corrected for a not perfect power substitution, i.e. $U_1 \neq U_2$, if any. If a good power control is achieved, this error propagates weakly on the final result and, in the experimental conditions which we are used to operate with, it can be neglected. Anyway, by virtue of the linearity of the sensor response, Eq. (7) has to be multiplied by $U_1/U_2$ to include the correction for imperfect power substitution.

IV. MODELS COMPARISON

Our analysis [11] consists in comparing the presented models against the effective efficiency of a lossless and perfectly adiabatic microcalorimeter:

$$\eta_e = e_2(e_1)^{-1}, \qquad (8)$$

Eq. (8) can be used to estimate the limiting accuracy obtainable in the power standard realization. This accuracy depends on the dc-measurement capability of the laboratory implementing the standard. Presently, for INRIM, the typical limit is of the order of parts in $10^4$ above 10 MHz, therefore no better result, in term of total accuracy, should be expected even after the complete error removal. The comparison has been carried out in terms of difference between the effective efficiencies evaluated with Eq. (6) and (7) (here named $\eta_6$ and $\eta_7$) and the limit value calculated with Eq. (8):

$$\begin{aligned} d_{1,2} &= \eta_{6,7} - \eta_e \\ U(d_{1,2}) &= \sqrt{U(\eta_{6,7})^2 - U(\eta_e)^2} \end{aligned} \qquad (9)$$

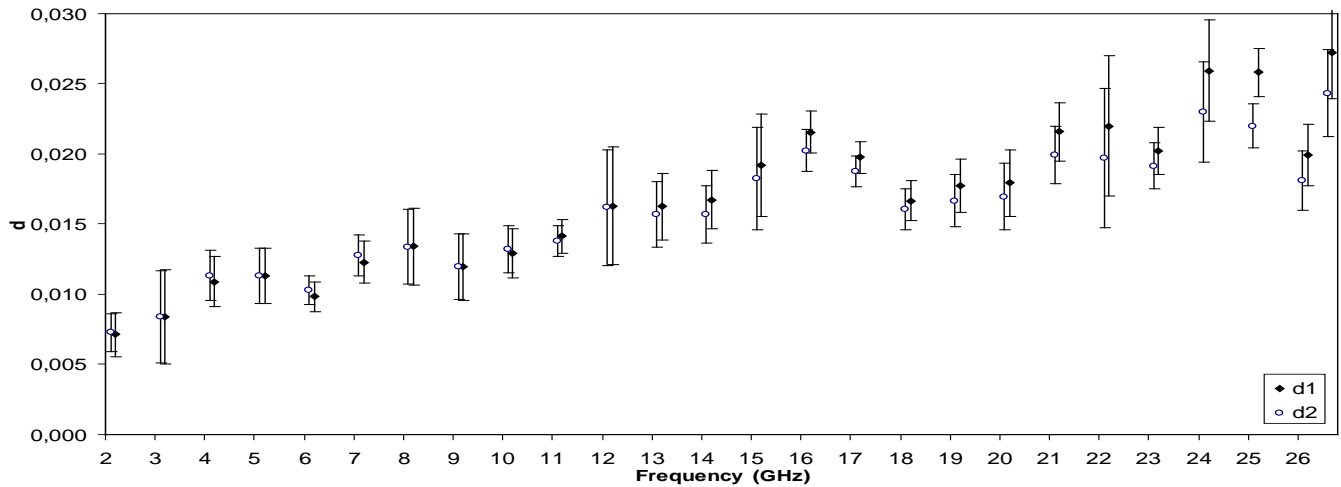

Fig. 3. Simultaneous plot of the difference $d_{1,2}$ between the effective efficiencies given by Eq. (6) and (7) and the ideal value (8). The results are relevant to a typical coaxial power standard realization in the 2-26.5GHz band (values are slightly x-axis shifted for best reading). At the highest frequencies all the corrections give significant contributions.

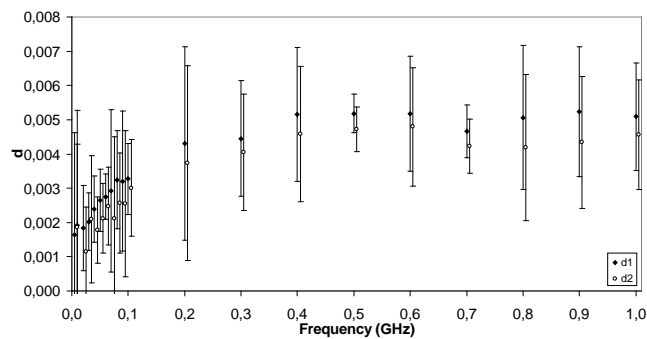

Fig. 4. Results obtained in a typical coaxial power standard realization in the band up to 1 GHz. Only the term relevant to feeding line losses is significant in this frequency range.

TABLE I
LIMITING AND MEASUREMENT UNCERTAINTY

| Freq. / GHz | Relative Limit Uncertainty | Relative Measurement Uncertainty |
|---|---|---|
| 0.01 | 0.21 | 0.27 |
| 0.1 | 0.07 | 0.12 |
| 1 | 0.11 | 0.12 |
| 10 | 0.12 | 0.13 |
| 20 | 0.19 | 0.20 |
| 26.5 | 0.26 | 0.29 |

Extended relative uncertainty ($k=2$) obtained applying the Law of Propagation of the Uncertainty to Eq. (8) (limit case) and to Eq. (7) (measurement values) for selected frequencies.

Figs. 3 and 4 show the results obtained in a realization of the coaxial power standard in the frequency band 10 MHz - 26.5 GHz by using a thermoelectric sensor fitted to 3.5 mm connector.

Squared points are relative to the effective efficiency corrected for the feeding line losses only (model (6)) while circles take also into account the losses in the short circuit used to calibrate the system (model (7)), respectively. It is evident, but also expected, that the figures obtained supposing an ideal microcalorimeter underestimate the effective efficiency. Moreover, the most significant correction is related to the HF losses on the feeding line. The correction related to the losses of the short-circuit implemented for the system calibration appears to have an effect only beyond 12 GHz in connection with an evident degradation of the short circuit reflection coefficient $\Gamma_{SC}$ (see Fig. 2). Indeed, at the highest frequency of the considered band, the two models differ of a not negligible quantity, so that the short circuit imperfections must be taken into account. Nevertheless, at low frequencies, this contribution is not important and the two effective efficiency evaluations are quite the same. The correction for the power substitution error has not been showed because, having in our case a good power control, it gives negligible contribution.

The total uncertainty associated with our coaxial microcalorimeter and, then, to our power standard realization, has been evaluated by using the standard Gaussian propagation on the mentioned models, according to the recommendation of the GUM [12]. The significant uncertainty components turned out to be the ones related to $e_1$, $e_2$, $e_{1SC}$, $e_{2SC}$ and $\Gamma_{SC}$. Thermo-voltages come from fitting procedures while the reflection coefficient from a Vector Network Analyzer evaluation. The sensor output voltages $U_1$ and $U_2$ are obtained from repeated measurements when their inclusion is necessary ($U_1 \neq U_2$).

Table 1 shows the total relative uncertainty for some selected frequencies in the band from 10 MHz to 26.5 GHz, reported with a coverage factor $k=2$. It can be seen that the uncertainty increases with frequency as expected but also at low frequency due to lack of sensitivity of the microcalorimeter. However, our measurement system shows a close approach to the limit uncertainty. In the whole range the results are well under 0.45%, though the majority of the points is below 0.25%.

V. CONCLUSIONS

We have performed a comparison among models suitable for describing a twin-line coaxial Microcalorimeter and its calibration process in the frequency range 10 MHz – 26.5

GHz.

The considered models are specifically related to the use of a transfer standard, or calorimetric load, of thermoelectric type, though they can be arranged also for bolometric power sensors.

The comparison shows that only the systematic error term related to the HF losses on the feeding line is really significant in all the considered frequency band. Independently of the model complexity, any other correction does not have a strong impact on the total accuracy of the power standard. Anyway, above a certain frequency (12 GHz in our case), the contribution of the reflecting standard used in the system calibration should be, in general, considered whereas the substitution error can be neglected if a proper power substitution is achieved.

We derived, in an original way, a model that accounts for realistic error sources. An aspect never adequately hint at by the literature has been also highlighted. It is the character of relative measurement of the effective efficiency. Indeed, an effective efficiency at the reference power appears that must be decided by the laboratory, because it cannot be measured.

If a new type of thermoelectric sensors will be available, this drawback could be removed. Such devices should have two independent heaters so to allow the HF/LF or HF/dc power substitution via separated transmission lines, as for bolometer detectors. This is enough to avoid the need of a common reference frequency and, furthermore, to remove the indetermination that arises in $U_1/U_2$, when the model is used to calibrate the microcalorimeter. Indeed when a short circuit is done on the HF feeding line, any sensor output is available and the $U_1/U_2$ ratio cannot be calculated, while with two independent heaters a power substitution may be operated so to have always a finite sensor output.

However Eq. (5) can be profitably used to check well the primary standard when the reference results of an international power comparison are available and to correct the same power standard for it, if any.

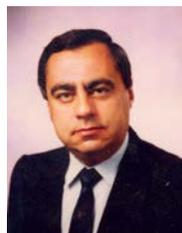
**Luciano Brunetti** was born in Asti, Italy, on September 11, 1951. He received the M.S. degree in Physics from the University of Torino, Italy, in 1977. Since 1977 he has been working at Istituto Nazionale di Ricerca Metrologica (INRIM, formerly IEN "Galileo Ferraris").

He has been dealing both with theoretical and experimental research in the field of high frequency primary metrology. His main task has always been the realization and the dissemination of the national standard of power, impedance and attenuation in the microwave range. In the last years he has been involved in the design and characterization of millimetre and microwave devices working at cryogenic temperature, and he collaborates also at the characterization of complex magnetic alloys at high frequency. Actually he is taking care of the extension of the national electrical standards in the millimetre wavelength range.

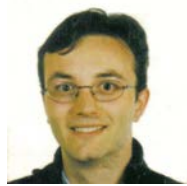
**Luca Oberto** was born in Pinerolo (Torino), Italy, on June 9, 1975. He received the M.S. degree in Physics from the University of Torino in 2003 and the PhD in Metrology from the Politecnico di Torino in 2008. From 2002 to 2003 he was with the Istituto Nazionale di Fisica Nucleare (INFN), Torino Section, working at the COMPASS experiment at CERN, Geneva, Switzerland. From 2003 he is with the Istituto Nazionale di Ricerca Metrologica (INRIM), Torino, Italy.

His research interests are in the field of high frequency metrology and in the realization and characterization of SIS mixers for astrophysical applications in the millimetre and sub-millimetre wave domain.

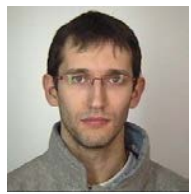
**Marco Sellone** was born in Torino, Italy, in 1979. He received the M.S. degree in Physics from the university of Torino in 2004 with a thesis on the CMS Electromagnetic Calorimeter of the CERN LHC accelerator. From 2005 he is working at the High Frequency Laboratory of the INRIM Electromagnetism Division – Istituto Nazionale di Ricerca Metrologica (formerly IEN "Galileo Ferraris") in Torino, and from 2006 he is a PhD student of the Politecnico di Torino.

His field of interest is the High Frequency Metrology mainly regarding AC-DC Transfer Difference measurements and High Frequency power standards.

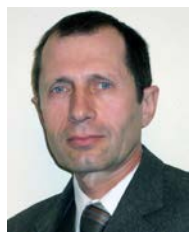
**Emil T. Vremera** was born in Neamt, Romania, on September 11, 1953. He received his degree in Electronics and the PhD in Electrical Measurements from the "Gh. Asachi" Technical University, Iasi, Romania in 1977 and 1998, respectively. In 1984, he joined the Department of Electric Measurements, Faculty of Electrical Engineering, Technical University of Iasi, first as Assistant Professor and then as Professor. He teaches electric and electronic measurements for the students in the electronic area.

Since 2001 he has been developing a research activity in RF Power measurements at Istituto Nazionale di Ricerca Metrologica (INRIM) Torino, Italy, formerly IEN "Galileo Ferraris". His main research interests concern measurement techniques of the electric and magnetic quantities, analogue to digital conversion for second-order quantities and virtual instrumentation.